\documentstyle[12pt]{article}

\begin{document}
\setcounter{page}{0}

\newcommand{\R}{{\rm I\!R}}
\newcommand{\Z}{{\sf Z\!\!Z}}
\newcommand{\goto}{\rightarrow}
\newcommand{\Goto}{\Rightarrow}

\newcommand{\al}{\alpha}
\renewcommand{\b}{\beta}
\renewcommand{\c}{\chi}
\renewcommand{\d}{\delta}
\newcommand{\D}{\Delta}
\newcommand{\ve}{\varepsilon}
\newcommand{\f}{\phi}
\newcommand{\F}{\Phi}
\newcommand{\vf}{\varphi}
\newcommand{\g}{\gamma}
\newcommand{\G}{\Gamma}
\newcommand{\k}{\kappa}
\renewcommand{\l}{\lambda}
\renewcommand{\L}{\Lambda}
\newcommand{\m}{\mu}
\newcommand{\n}{\nu}
\newcommand{\r}{\rho}
\newcommand{\vr}{\varrho}
\renewcommand{\o}{\omega}
\renewcommand{\O}{\Omega}
\newcommand{\p}{\psi}
\renewcommand{\P}{\Psi}
\newcommand{\s}{\sigma}
\renewcommand{\S}{\Sigma}
\newcommand{\th}{\theta}
\newcommand{\vt}{\vartheta}
\renewcommand{\t}{\tau}
\newcommand{\vp}{\varphi}
\newcommand{\x}{\xi}
\newcommand{\z}{\zeta}
\newcommand{\ta}{\triangle}
\newcommand{\w}{\wedge}
\newcommand{\e}{\eta}
\newcommand{\Th}{\Theta}
\newcommand{\td}{\tilde}
\newcommand{\vep}{\varepsilon}
\newcommand{\ep}{\epsilon}
\newcommand{\na}{\nabla}
\newcommand{\be}{\begin{equation}}
\newcommand{\ee}[1]{\label{#1}\end{equation}}
\newcommand{\bE}{\begin{eqnarray}}
\newcommand{\eE}[1]{\label{#1}\end{eqnarray}}
\newcommand{\nn}{\nonumber}
\renewcommand{\thefootnote}{\fnsymbol{footnote}}

\newcommand{\CAG}{{\cal A/\cal G}}
\newcommand{\CA}{{\cal A}}
\newcommand{\CB}{{\cal B}}
\newcommand{\CC}{{\cal C}}
\newcommand{\CD}{{\cal D}}
\newcommand{\CF}{{\cal F}}
\newcommand{\CG}{{\cal G}}
\newcommand{\CH}{{\cal H}}
\newcommand{\CI}{{\cal I}}
\newcommand{\CL}{{\cal L}}
\newcommand{\CO}{{\cal O}}
\newcommand{\CP}{{\cal P}}
\newcommand{\CQ}{{\cal Q}}
\newcommand{\CR}{{\cal R}}
\newcommand{\CT}{{\cal T}}
\newcommand{\CU}{{\cal U}}
\newcommand{\CW}{{\cal W}}
\newcommand{\CM}{{\cal M}}
\newcommand{\CS}{{\cal S}}
\newcommand{\CN}{{\cal N}}

\newcommand{\siml}{\raisebox{-.6ex}{$\stackrel{<}{\displaystyle{\sim}}$}}
\newcommand{\simg}{\raisebox{-.6ex}{$\stackrel{>}{\displaystyle{\sim}}$}}
\newcommand{\ind}{\scriptscriptstyle}

\newcommand{\rd}{\partial}
\newcommand{\grad}[1]{\,\nabla\!_{{#1}}\,}
\newcommand{\gradd}[2]{\,\nabla\!_{{#1}}\nabla\!_{{#2}}\,}
\newcommand{\om}[2]{\omega^{#1}{}_{#2}}
\newcommand{\vev}[1]{\langle #1 \rangle}
\newcommand{\lrarr}{\longrightarrow}
\newcommand{\darr}[1]{\raise1.5ex\hbox{$\leftrightarrow$}\mkern-16.5mu #1}
\newcommand{\Ha}{{\frac{1}{2}}}
\newcommand{\ha}{{\textstyle{\frac{1}{2}}}}
\newcommand{\fr}[2]{{\textstyle{#1\over#2}}}
\newcommand{\Fr}[2]{{#1 \over #2}}
\newcommand{\dt}{{\frac{d}{dt}}}
\newcommand{\rf}[1]{\fr{\rd}{\rd #1}}
\newcommand{\rF}[1]{\Fr{\rd}{\rd #1}}
\newcommand{\Rf}[2]{\fr{\rd #1}{\rd #2}}
\newcommand{\RF}[2]{\Fr{\rd #1}{\rd #2}}
\newcommand{\df}[1]{\fr{\d}{\d #1}}
\newcommand{\dF}[1]{\Fr{\d}{\d #1}}
\newcommand{\DF}[2]{\Fr{\d #1}{\d #2}}
\newcommand{\DDF}[3]{\Fr{\d^2 #1}{\d #2\d #3}}
\newcommand{\ddf}[2]{\Fr{d^2 #1}{d #2^2}}
\newcommand{\DDDF}[4]{\Fr{\d^3 #1}{\d #2\d #3\d #4}}
\newcommand{\ddF}[3]{\Fr{\d^n#1}{\d#2\cdots\d#3}}
\newcommand{\fs}[1]{#1\!\!\!/\,}   
\newcommand{\Fs}[1]{#1\!\!\!\!/\,} 
\newcommand{\pa}{p_a \dot{x}^a}
\newcommand{\slag}{\Fr{i}{4}\l ^{ab}\,Tr\,\s_{ab}\,\L^{-1}\dot{\L}}
\newcommand{\lag}{\CL\,=p_a \dot{z}^a+\Fr{i}{2}\l ^{ab}\,Tr,s_{ab}\
\L^{-1}\dot{\L}}
\newcommand{\spin}{\Fr{1}{2}\l^{ab}\L\,\s_{ab}\,\L^{-1}}
\newcommand{\Df}[1]{\Fr{d}{d #1}}
\newcommand{\hlkl}{\Fr{\l^{kl}}{2}}
\newcommand{\stcon}{C_{ab,cd}{}{}{}^{ef}}

\newcommand{\cmp}[3]{, Comm.\ Math.\ Phys.\ {{\bf #1}} {(#2)} {#3}.}
\newcommand{\pl}[3]{, Phys.\ Lett.\ {{\bf #1}} {(#2)} {#3}.}
\newcommand{\np}[3]{, Nucl.\ Phys.\ {{\bf #1}} {(#2)} {#3}.}
\newcommand{\pr}[3]{, Phys.\ Rev.\ {{\bf #1}} {(#2)} {#3}.}
\newcommand{\prl}[3]{, Phys.\ Rev.\ Lett.\ {{\bf #1}} {(#2)} {#3}.}
\newcommand{\ijmp}[3]{, Int.\ J.\ Mod.\ Phys.\ {{\bf #1}} {(#2)} {#3}.}
\newcommand{\mpl}[3]{, Mod.\ Phys.\ Lett.\ {{\bf #1}} {(#2)} {#3}.}
\newcommand{\jdg}[3]{, J.\ Diff.\ Geo.\ {{\bf #1}} {(#2)} {#3}.}
\newcommand{\pnas}[3]{, Proc.\ Nat.\ Acad.\ Sci.\ USA.\ {{\bf #1}} {(#2)}
{#3}.}
\newcommand{\zp}[3]{, Z.\ Phys.\ {{\bf #1}} {(#2)} {#3}.}
\newcommand{\ap}[3]{, Ann.\ Phys.\ {{\bf #1}} {(#2)} {#3}.}
\newcommand{\jmp}[3]{, J.\ Math.\ Phys.\ {{\bf #1}} {(#2)} {#3}.}
\newcommand{\nc}[3]{, Il\ Nuovo\ Cimento\ {{\bf #1}} {(#2)} {#3}.}
\newcommand{\prs}[3]{, Proc.\ Roy.\ Soc.(London)\ {{\bf #1}} {(#2)} {#3}.}

\newcommand{\PR}{Phys. Rev. }
\newcommand{\PRL}{Phys. Rev. Lett. }
\newcommand{\NP}{Nucl. Phys. }

\newpage
\setcounter{page}{0}
\begin{titlepage}
\begin{flushright}
\hfill{KAIST-CHEP-93/10}\\
\vskip 2mm
\hfill{YUMS-93-09}
\end{flushright}
\vspace{0.7cm}
\begin{center}
{\Large\bf A Covariant Formulation}
\vskip 3mm
{\Large\bf of Classical Spinning Particle}
\vskip 1.5cm
{\bf Jin-Ho Cho \footnote{e-mail address: jhcho@chiak.kaist.ac.kr}$^1$,\,}
{\bf Seungjoon Hyun \footnote{e-mail address:
hyun@phya.yonsei.ac.kr}$^2$\,\,} and
{\bf Jae-Kwan Kim}$^1$
\vskip 0.3cm
$^1${\sl Department of Physics\\
Korea Advanced Institute of Science and Technology\\
373-1 Yusung-ku, Taejon, 305-701, Korea}
\vskip 0.3cm
$^2${\sl Institute for Mathematical Sciences\\
Yonsei University, Seoul, 120-749, Korea}
\vskip 0.5cm
\end{center}

\setcounter{footnote}{0}

\begin{abstract}
\noindent
Covariantly we reformulate the description of a spinning particle in terms
of the Poincar\'{e} group. We also construct a Lagrangian
which entails all possible constraints explicitly; all constraints can
be obtained just from the Lagrangian. Furthermore, in this covariant
reformulation, the Lorentz element is to be considered to evolve the
momentum or spin component from an arbitrary fixed frame and not just
from the particle rest frame. In distinction with the usual formulation,
our system is directly comparable with the pseudo-classical formulation.
We get a peculiar symmetry which resembles the supersymmetry of the
pseudo-classical formulation.
\end{abstract}

\end{titlepage}

\newpage
\renewcommand{\thefootnote}{\arabic{footnote}}
Inspired by the spinning string, many studies have paid attentions to the
spinning particle. There are two standard ways of describing the
spin degrees of freedom; the `classical' way describes them in terms of
the Lorentz group elements \cite{rha}
and the `pseudo-classical' one does that in terms of the Grassmann
quantities \cite{rbe}\cite{rbr}.

In the present paper we shall be concerned about the classical formulation
in such a way that enlightens the relation with the pseudo-classical
description. We shall begin with the `classical' description
and reformulate it in a form comparable with the `pseudo-classical'
one; in the covariant fashion and without extra constraints assumed out
of the Lagrangian. Further we shall investigate the symmetry analogous
to that supersymmetry which the `pseudo-classical' formulation entails.

The `classical' description uses the Poincar\'{e} group elements
$(\L^a{}_b,~q^c),~~\L~\in SO(3,1)^\uparrow_{+}$ (hereafter just denoted
by $SO(3,1)$,) as the dynamical variables \cite{rba}.
The Poincar\'{e} transformations for those elements are given by the
standard multiplication;
$(\tilde{\L},~\tilde{a})~(\L(t),~q(t))
=(\tilde{\L}~\L(t),~\tilde{\L}~q(t)+\tilde{a}).$ We can find two invariant
one forms $~\L^{-1}~d\L~$ and $~\L^{-1}~dq$ \cite{rha}, which constitute
the Poincar\'{e} invariant Lagrangian
\be
\CL_{cl}\,=p^a \dot{q}_a+\Fr{i}{4}\l ~\,Tr\,\s_{12}\
\L^{-1}\dot{\L},
\ee{ncl}
where
\be
p^a\equiv m~\L^a{}_0,\,\,(~m~>~0)~
\ee{emm}
is the momentum, $q_a$ are interpreted as the coordinates of the
particle and $\l\,\s_{12}$ is an element in $so(3,1)$.\footnote{notation
convention : without extra specification, Latin indices run from $0$ to $3$
and raised or lowered by the Minkowski metric $\e=\{-1,1,1,1\}$ and
repeated indices will be summed over. The dot over
the variables means the derivative with respect to the parameter $t$.}
The first term is the usual kinetic term $-m~\sqrt{-(\L^{-1}\dot{q})^2}$
written in the first order style with the above definition (constraint)
(\ref{emm}) for the momentum $p^a$. The second term induces local $U(1)$
symmetry $\L~\goto~\L~e^{i\al\s_{12}},$ with the fixed (stabilized) element
$\Ha\,S^{ab}\,\s_{ab}\,\equiv\,\Ha\l~\L~\s_{12}~\L^{-1}~\in~so(3,1),$
which define `the spin component' $S^{ab}.$
With the representation
$(\s_{ab})^{cd}=\,-i\,(\d^c_a\,\d^d_b\,-\d^d_a\,\d^c_b)$, The spin
component $S^{ab}$ is rewritten in the form
\be
S^{ab}\,=\,\l(\L^a{}_1\,\L^b{}_2-\L^a{}_2\,\L^b{}_1).
\ee{spc}
Therefore the Lorentz elements $\L$ describe the spin degrees
of freedom while they also define the momentum through the transformation
of mass from {\it the particle rest frame} (see (\ref{emm})).
However, the element $\l\,\s_{12}$ is non-covariantly chosen in the
Lorentz algebra, so that $\l~\s_{12}$ is interpreted as `the spin
momentum' in {\it the particle rest frame}.
\footnote{In the particle rest frame,
the $p^a=m\d^a{}_0$ and
$S_{ab}=\Ha\l(\d_a{}^1\d_b{}^2-\d_b{}^1\d_a{}^2)$ and they
satisfy the relation $p^a~S_{ab}=0$. The spin component $S_{ab}$
satisfying $p^aS_{ab}=0$ will be called `the spin momentum'; it corresponds
to the non-relativistic spin vector.}
Moreover, formally the system appears to have no relation with the
`pseudo-classical' system where two constraints are involved
\cite{rgal} and those
two constraints become the Klein-Gordon equation and the Dirac equation
respectively upon quantization.\footnote{in the `classical' system above,
no need for the extra constraint $p^2+m^2=0$ corresponding to the
Klein-Gordon equation, because the momentum is already
defined to satisfy that.}

In this work, we reformulate the `classical' spinning
particle in the covariant fashion and construct a Lagrangian which
entails all possible constraints explicitly. In this covariant formulation,
the Lorentz element is to be considered to evolve the momentum or spin
component from {\it an arbitrary fixed frame} and not just from the
particle rest frame. We analyze the constraint structure
{\it\`{a} la} Dirac and obtain three first class
constraints. While one of them is concerned with the reparametrization,
the other two are unusual symmetries similar to
the supersymmetry of the pseudo-classical formulation.
We obtain `the physical spin momentum'
and from the
Pauli-Lubanski vector, get the spin value available.


We start from the first order Lagrangian
\be
\CL_{cl}\,=p^a \dot{x}_a-\Fr{\l^{kl}}{2}~t_k\dot{t}_l
-\Fr{\l^{kl}}{2}~\L_{ck}\dot{\L}^c{}_l-M_k\l^{kl}(p_b\L^b{}_l+m~t_l)
-N(p^2+m^2).
\ee{ela}
The third term may be rewritten as
$-\Fr{\l^{kl}}{2}~\L_{ck}\dot{\L}^c{}_l=\Fr{i}{4}\l ^{kl}\,Tr\,\s_{kl}\
\L^{-1}\dot{\L}$, i.e., it is Maura-Cartan one form of $SO(3,1)$ projected
on the specific direction $\Ha \l^{kl}\s_{kl}~\in~so(3,1).$
For covariance, $\l^{kl}$ is assumed
to be invertible.\footnote{For degenerate $\l^{kl}$, we can work
on the reduced phase space and expect to have the same result with this
paper. However for covariant formulation, we will not do so here.}
It should be noticed that in this formulation $\l^{0i}\ne 0,~i=1,2,3$,
thus $\l^{kl}$ can't be the spin momentum in {\it the particle rest
frame.} Thus in the covariant definition of spin component
$S^{ab}=\L^a{}_c\L^b{}_d\l^{cd},$ $\L$ is no longer the transformation
from {\it the particle rest frame}. The true interpretation of
$\L$ is given by the constraint term $-M_k\l^{kl}(p_b\L^b{}_l+m~t_l),$
where $M_k$ are the Lagrange multipliers. The term tells that the momentum
$p^a$ is to be obtained by the Lorentz transformation $\L$
from {\it the fixed frame} where the momentum is given by
$-m~t^a$; $p^a=-m~\L^a{}_b~t^b.$ And $t^b\equiv \bar{\L}^{b0}$ is
the Lorentz element which transforms the particle momentum from the
particle rest frame to the fixed frame mentioned above.\footnote{$t^k$
is invariant under the Poincar\'{e} transformation
to retain the relation $p^a=-m~\L^a{}_b~t^b.~$;
$t^b\equiv \bar{\L}^{b0}=\L^{-1bc}\tilde{\L}_c{}^0,$ where $\tilde{\L}$ is
the transformation from the particle rest frame to the observer frame.}

The system has at least two symmetries; global Poincar\'{e} symmetry and
local reparametrization symmetry. Those symmetries are essential to the
relativistic description of a spinning particle, as is the case with a
spinless particle. We note that for $\l^{kl}\to 0$, we can recover
the usual relativistic description of a spinless particle.

Since the Lagrangian is in the first order, we can directly read off
the Poisson brackets from the dynamical terms \footnote{For the
Lagrangian of the form $\Ha q^a\o_{ab}\dot{q}^b-H(q)$ with
non-degenerate $\o_{ab}$, the equation of
motion is given by $\dot{q}^a=\o^{-1ab}\rf{q^b}H(q),$ which is
rewritten as $\dot{q}^a=\{q^a,H(q)\}$ with the
Poisson bracket defined as $\{q^a,q^b\}=\o^{-1ab}$ \cite{rja}.}
\be
\{~x^a,~p^b~\}=\e^{ab},~~\{~t_k,~t_l~\}=-\l^{-1}_{kl}.
\ee{epb}
As for $\L^a{}_b$, we should be careful because they are in fact group
elements so constrained by another condition
$\L_{ck}\L^c{}_l-\e_{kl}=0$. To get the Poisson bracket for these
group elements we will follow the method in \cite{rba}. $\L\in SO(3,1)$ is
parametrized by $\x=(\x_1,\x_2,\cdots,\x_6).$
The canonical brackets for $\x_a$ and their conjugate momenta $\pi^b$ are
given by
\be
\{~\x_\al,~\x_\b~\}=\{~\pi_\al,~\pi_\b~\}=0,
\ee{expi}
For convenience, we replace those phase space variables with the ones
that are easier to treat. Prior to that,
a basic identity should be noted.
We define a set of functions $f_i(\vep),i=1,\cdots,6$:
\be
e^{\ha\s_{ab}\vep^{ab}}\L(\x)=\L(f(\vep)),~f(0)=\x,
\ee{ext}
where $\s_{ab}$
are the group generators of $so(3,1)$ satisfying
\be
[\s_{ab},\s_{cd}]=\Fr{i}{2}\stcon\s_{ef}.
\ee{eal}
Differentiation (\ref{ext}) with respect to $\vep$ at $\vep =0$ gives
\be
i\s_{ab}\L(\x)=
{\RF{f_\al}{\vep^{ab}}}\vert_{\vep=0}~{\RF{\L}{f_\al}}\vert_{\vep=0}
\equiv N_{\al,ab}\RF{\L}{\x_\al},
\ee{edi}
where $det~N \ne 0$ because, otherwise,
$\s_{ab}$ will not be linearly independent one another \cite{rba}.

Since $det~N \ne 0,$ we can transform the canonical momenta $\pi^a$ to
get new variables
\be
t_{ab}=-\pi^\al~N_{\al,ab}=-\RF{\CL}{\dot{\x}_\al}N_{\al,ab},
\ee{ett}
which satisfy
\bE
&&\{~t_{ab},~\L~\}=i\s_{ab}\L,\label{etla}\\
&&\{~t_{ab},~\L^{-1}~\}=-i\L^{-1}\s_{ab},\label{etlb}\\
&&\{~t_{ab},~t_{cd}~\}=\stcon t_{ef}.
\eE{etl}
To obtain (\ref{etl}) the Jacobi identity was used \cite{rba}.
With the Lagrangian (\ref{ela}), $t_{ab}$ can be explicitly rewritten as
\bE
t_{ab}&=&\hlkl \L_{ck}\RF{\L^c{}_l}{\x_\al}N_{\al,ab}
=\l^{kl}\L_{ak}\L_{bl}\nn\\
&=&S_{ab}.
\eE{eti}
This is just the primary constraint concerned with the first order
dynamical term $-\hlkl \L_{ck}\dot{\L}^c{}_l$.


Here we analyze the Hamiltonian structure \`{a} la Dirac to look for full
gauge symmetries \cite{rDi}.
The total Hamiltonian consists of only the primary constraints:
\bE
\CH_T&=&N(p^2+m^2)+M_k\l^{kl}(p_a\L^a{}_l+m~t_l)+\Ha
L_{ab}(t^{ab}-\L^a{}_c\L^b{}_d\l^{cd})\nn\\
&\equiv&N\al+M_k\l^{kl}\b_l+\Ha L_{ab}\g^{ab}.
\eE{eham}
It is easy to get the non-vanishing Poisson brackets for those primary
constraints;
\bE
&&\{~\b_k,~\b_l~\}=-m^2\l^{-1}_{kl},\nn\\
&&\{~\b_k,~\g_{ab}~\}=p_b\L_{ak}-p_a\L_{bk},\nn\\
&&\{~\g_{ab},~\g_{cd}~\}=\stcon (t_{ef}-2\L_{eg}\L_{fh}\l^{gh}).
\eE{abc}

Now we require those constraints to be constants of
motion for this total Hamiltonian.
\bE
&&\dot{\al}=0,\label{econsa}\\
&&\dot{\b_k}=m^2M_k+\Ha L_{ab}(p^b\L^a{}_k-p^a\L^b{}_k)\approx
0,\label{econsb}\\
&&\dot{\g}_{ab}=M_k\l^{kl}(p_a\L_{bl}-p_b\L_{al})
+\Ha L^{cd}\stcon (t_{ef}-2\L_{eg}\L_{fh}\l^{gh})\approx 0.
\eE{econs}
Therefore no secondary constraint arises. From these equations, we can
determine the Lagrange multipliers. From (\ref{econsb},\ref{econs}), we get
\be
L^{cd}(\CM_{ac}S_{bd}-\CM_{bc}S_{ad})=0,
\ee{eaa}
where $\CM_{ab}\equiv\e_{ab}+\Fr{p_ap_b}{m^2}$ is the
projection operator onto the direction normal to $p^a$.
The solution is easily found as
$L^{ab}=S^{-1~ab}$ or $L^{ab}=\CN^a{}_cS^{cb}+S^{ac}\CN_c{}^b,$
where
$\CN^a{}_c\equiv -\Fr{p^ap_c}{m^2}$ is the projection operator onto the
direction along $p^a.$

Hence with $L^{ab}=L S^{-1~ab}+M(\CN^a{}_cS^{cb}+S^{ac}\CN_c{}^b),$
the total Hamiltonian can be rewritten as follows.
\bE
\CH_T&=&N\al+M\Fr{p_cS^c{}_d\L^d{}_k}{m^2}
(\L_a{}^kt^{ab}p_b+m\l^{ka}t_a)
+L(\Fr{p_a\L^{ab}t_b}{m^2}+\Ha S^{-1}_{ab}t^{ab}+\Fr{p^2}{m^2}+2)\nn\\
&\equiv&N\al+M\tilde{\b}+L\tilde{\g},
\eE{eth}
where $N,~M$ and $L$ are undetermined multipliers.

Since those three Lagrange multipliers $N,~M$ and $L$ in $\CH_T$ are
undetermined, their corresponding constraints
$\al,~\tilde{\b}$ and $\tilde{\g}$ are first class. Further their Poisson
algebra can be shown to be trivial
by simple calculation; all the Poisson brackets vanish strongly.
Being first class, $\al,~\tilde{\b}$ and $\tilde{\g}$ are concerned with
local symmetries and generate the following transformations,
\bE
&&\{x^a,\al\}^*=2p^a,\nn\\
&&\{p^a,\al\}^*=\{t^a,\al\}^*=\{\L^a{}_b,\al\}^*=0,
\eE{esya}

\bE
&&\{x^a,\tilde{\b}\}^*=\Fr{S^{ab}S_{bc}p^c}{m^2},\nn\\
&&\{p^a,\tilde{\b}\}^*=0,\nn\\
&&\{t^a,\tilde{\b}\}^*=\Fr{p^cS_{cb}\L^{ba}}{m},\nn\\
&&\{\L^a{}_b,\tilde{\b}\}^*=\Fr{1}{m^2}(S^{ac}p_cp_d\L^d{}_b
+p^ap^cS_{cd}\L^d{}_b),
\eE{esyb}
and

\bE
&&\{x^a,\tilde{\g}\}^*=\Fr{\L^{ab}t_b}{m}+\Fr{2p^a}{m^2},\nn\\
&&\{p^a,\tilde{\g}\}^*=0,\nn\\
&&\{t^a,\tilde{\g}\}^*=\Fr{p^c\L_{cb}\l^{-1ba}}{m},\nn\\
&&\{\L^a{}_b,\tilde{\g}\}^*=-S^{-1a}{}_c\L^c{}_b,
\eE{esyg}
where $\{\cdot,\cdot\}^*$ is Dirac bracket, which is equal to the
corresponding Poisson bracket because $\al,~\tilde{\b}$ and $\tilde{\g}$
are first class.


Now we are ready to get the equations of motion.\footnote{also may
be obtained through the variational principle.}
\bE
\dot{p}^a&=&\{~p^a,~\CH_T\}=0,\label{eoma}\\
\dot{x}^a&=&\{~x^a,~\CH_T\}
=2Np^a+\Fr{1}{m^2}(Lp^a+MS^{ab}S_{bc}p^c),\label{eomb}\\
\dot{t}^a&=&\{~t^a,~\CH_T\}
=\Fr{1}{m}(Mp^cS_{cb}\L_{ba}+Lp^c\L_{cb}\l^{-1ba}),\label{eomc}\\
\dot{\L}^a{}_b&=&\{~\L^a{}_b,~\CH_T\}=\Fr{M}{m^2}(S^{ac}p_cp_d\L^d{}_b+p^a
p^cS_{cd}\L^d{}_b)-LS^{-1a}{}_c\L^c{}_b.
\eE{eom}
The equation (\ref{eom}) can be rewritten in more familiar form;
\bE
\dot{S}^{ab}&=&\dot{\L}^a{}_c\L^b{}_d\l^{cd}-(a\leftrightarrow b)
=\Fr{M}{m^2}p^kS_{kl}(p^aS^{lb}-p^bS^{la})\nn\\
&=&-\dot{x}^ap^b+\dot{x}^bp^a\nn\\
&\Rightarrow&\Df{t}(x_ap_b-x_bp_a+S_{ab})=0.
\eE{escon}
Therefore (\ref{eoma}) and (\ref{escon}) tell the conservation of $p^a$ and
$J^{ab}\equiv x^ap^b-x^bp^a+S^{ab}$. We can check the symmetry concerned
with these conserved quantities through the Dirac algebra (the algebra with
the Dirac bracket\cite{rDi} as its binary operation) of them because the
algebra becomes the Lie algebra upon quantization.

However in calculating the Dirac bracket, we have a practical
difficulty of extracting the second class constraints {\it in a covariant
way}.
Thus we sidestep this difficulty by substituting `corresponding first class
variables' for those conserved charges.
Since all Dirac brackets involving
first class are equal to the corresponding Poisson brackets, we are able
to get the correct Dirac algebra just through the Poisson algebra
\cite{rba}. While $p^a$ is already first class, $J^{ab}$ is not.
\be
\{J^{ab},\b^k\}=p^b\L^{ak}-p^a\L^{ak}\ne 0.
\ee{ejj}
We construct the following first class variable which is `weakly'
\cite{rha}
\cite{rDi} equal to $J^{ab}.$
\be
J^{ab*}=J^{ab}+\g^{ab}=x^ap^b-x^bp^a+t^{ab}.
\ee{ejjj}
Now the Dirac brackets for these variables are given by
\bE
&&\{p^a,\cdot\}^*=\{p^a,\cdot\}\nn\\
&&\{J^{ab},\cdot\}^*\equiv \{J^{ab*},\cdot\},
\eE{edd}
where $\{\cdot,\cdot\}^*$ denotes the Dirac bracket. Hence we easily get
the following Dirac algebra.
\bE
&&\{p^a,J^{cd}\}^*=\{p^a,J^{cd}\}=-\e^{ac}p^d+\e^{ad}p^c\nn\\
&&\{J_{ab},J_{cd}\}^*\equiv \{J_{ab}{}^*,J_{cd}{}^*\}
=\stcon J_{ef}{}^*\approx\stcon J_{ef}.
\eE{edal}
This algebra is just the Poincar\'{e} algebra and thus the conserved
quantities
$p^a$ and $J^{ab}$ are the momentum and the angular momentum respectively ;
The interpretation of $p^a$ and $S^{ab}$ (as the momentum and the spin
component respectively) is justified.

Before concluding this section we give some subsidiary notes.
First, the momentum and angular momentum are observables. That is, they
satisfy
\bE
&&\{p^a,\al\}=\{p^a,\tilde{\b}\}=\{p^a,\tilde{\g}\}=0\nn\\
&&\{J^{ab},\al\}=\{J^{ab},\tilde{\b}\}=\{J^{ab},\tilde{\g}\}=0.
\eE{eobs}
Second, from (\ref{eomb}) and (\ref{eomc}) we get
\be
\dot{x}^a=2Np^a+\Fr{1}{m}\L^{ak}\l_{lk}\dot{t}^l
=2Np^a+\Fr{1}{m}S^{ka}\L_{kl}\dot{t}^l,
\ee{etxp}
which tells that the momentum $p^a$ is no
longer parallel with $\dot{x}^a$ due to that last term which is the
classical analogue of {\it Zitterbewegung} \cite{fDD}.
Third, making use of (\ref{etxp}) and $p^2+m^2=0$, we can
eliminate $p,~N$ and $M$ to obtain the usual second order Lagrangian
\be
\CL=-m\sqrt{-(\dot{x}^a-\Fr{1}{m}\L^{ak}\l_{lk}\dot{t}^l)^2}
+\hlkl t_k\dot{t}_l -\hlkl \L_{ck}\dot{\L}^c{}_l.
\ee{esec}


As mentioned earlier, $S^{ab}$ is not `the spin momentum' because
$S^{ab}p_b\ne 0.$  Moreover it
is not a physical observable. That is, it does not commute with all the
first class constraints:
\be
\{S^{ab},\tilde{\b}\}=\Fr{1}{m^2}(-S_a{}^dS_{dc}p^cp_b
+S_b{}^dS_{dc}p^cp_a).
\ee{ephy}
In this section we are to find {\it the physical spin
momentum} and determine the spin value available for this spinning
particle.

We project the conserved angular momentum $J^{ab}$ on the
direction normal to the momentum $p^a.$
\bE
\tilde{S}^{ab}&\equiv&\CM^a{}_c\CM^b{}_dJ^{cd}\nn\\
&=&S^{ab}+\Fr{1}{m^2}(p^ap_cS^{cb}-p^bp_cS^{ca}),
\eE{epro}
then it is obvious that
$\tilde{S}^{ab}p_b=0.$ Further so defined $\tilde{S}^{ab}$ is a physical
observable because
\be
\{\tilde{S}^{ab},\al\}=\{\tilde{S}^{ab},\tilde{\b}\}
=\{\tilde{S}^{ab},\tilde{\g}\}=0.
\ee{stil}
Hence $\tilde{S}^{ab}$ can be thought of as `the physical spin momentum'.
This assertion can be justified by checking
the Dirac algebra. We again replace that physical spin momentum with the
corresponding first class variable :
\be
\tilde{S}^{ab*}=\CM^a{}_c\CM^b{}_dJ^{cd}.
\ee{esfi}
Then the Dirac bracket for $\tilde{S}^{ab}$ is given by the Poisson bracket
for $\tilde{S}^{ab*}$ :
\bE
\{\tilde{S}^{ab},\tilde{S}^{cd}\}^*&
=&\{\tilde{S}^{ab*},\tilde{S}^{cd*}\}\nn\\
&=&\CM^{ac}\tilde{S}^{bd}-\CM^{bc}\tilde{S}^{ad}
+\CM^{bd}\tilde{S}^{ac}-\CM^{ad}\tilde{S}^{bc}.
\eE{esta}
This seemingly peculiar algebra becomes rather a familiar one in the
particle rest frame, where $\tilde{S}^{a0}=0,~p^i=0$ and
$\tilde{S}^{ij}=S^{ij}$:
\be
\{\tilde{S}^{ij},\tilde{S}^{kl}\}^*=\e^{ik}\tilde{S}^{jl}
-\e^{jk}\tilde{S}^{il}+\e^{jl}\tilde{S}^{ik}-\e^{il}\tilde{S}^{jk}.
\ee{eij}
With $\bar{S}^{i}=\Ha \vep^{ijk}\tilde{S}_{jk}$, we obtain the well
known algebra for the non-relativistic spin vector,
\be
\{\bar{S}^i,\bar{S}^j\}^*=\vep^{ijk}\bar{S}_k.
\ee{esvec}

Finally, we get the spin value for the above physical spin momentum.
 From the Pauli-Lubanski vector $W^a=\vep^{abcd}J_{bc}p_d,$
we obtain
\bE
W^2&=&-\Ha S_{ab}S^{ab}p^2+S_{ab}p^bS^{ac}p_c\nn\\
&=&-\Ha \tilde{S}_{ab}\tilde{S}^{ab}p^2=\Fr{m^2}{2}S_{ij}S^{ij}
\equiv \Fr{m^2}{2}\tilde{\l}_{ij}\tilde{\l}^{ij}.
\eE{eplk}
In the quantum theory, as the Dirac bracket is replaced with the
commutator, (\ref{esvec}) becomes
\be
[~\widehat{S}^i,\widehat{S}^j~]=\vep_{ijk}\widehat{S}_k,
\ee{esoii}
where
$\widehat{S}^i$ is the quantum operator corresponding to $\bar{S}^i$. It is
well known \cite{rsa} that this
$so(3)$ algebra determines the spin value as
\be
\Ha\tilde{\l}_{ij}\tilde{\l}^{ij}=l~(l+1),~~l=0,1/2,1,\cdots .
\ee{eige}


So far we have dealt with a modified classical formalism describing the
spinning particle. Based on the Poincar\'{e} invariance, the system has
two conserved quantities which proves to be the momentum and the
angular momentum. From the angular momentum we extracted the spin
momentum which actually gives arbitrary spin value. However at the
quantum level the available spin values are discretized to integers or
half integers. These results are similar to \cite{rba}.

However this system also has many things in common with the
pseudo-classical formalism \cite{rgal}.
First, the spin component hinders the momentum $p^a$ from being parallel
with $\dot{x}^a$ as we see in (\ref{etxp}) (the classical Zitterbewegung)
\cite{rbe}.
Moreover due to this fact, $\al$ does not generate mere
reparametrization. Indeed the transformation for $x^a$ is given by
\be
\d x^a=\{x^a,~\d t~\al\}=2~\d t~p^a
=\Fr{2m~\d t}{\sqrt{-z^2}}(\dot{x}^a-\Fr{1}{m}\L^{ak}\l_{lk}\dot{t}^l),
\ee{ezz}
where $z^a=\dot{x}^a-\Fr{1}{m}\L^{ak}\l_{lk}\dot{t}^l.$ Besides the
reparametrization, $\al$ generate the second term. This is not a peculiar
property of this system; it is common with the pseudo-classical
formulation.

Second, the system has extra first class constraints $\tilde{\b}$
and $\tilde{\g}$ other than the usual $\al$. Let us consider the following
transformation
\bE
\d x^a&=&\{~x^a,~M\tilde{\b}+L\tilde{\g}\}
=\Fr{1}{m^2}(Lp^a+MS^{ab}S_{bc}p^c),\label{ettta}\\
\d t^a&=&\{~t^a,~M\tilde{\b}+L\tilde{\g}\}
=\Fr{1}{m}(Mp^cS_{cb}\L_{ba}+Lp^c\L_{b}\l^{-1ba}),\label{etttb}\\
\d \L^a{}_b&=&\{~\L^a{}_b,~M\tilde{\b}+L\tilde{\g}\}
=\Fr{M}{m^2}(S^{ac}p_cp_d\L^d{}_b+p^ap^cS_{cd}\L^d{}_b)
-LS^{-1a}{}_c\L^c{}_b.
\eE{ettt}
With $M_k=\Fr{\L^b{}_k}{m^2}(Lp^aS^{-1}_{ab}+Mp^aS_{ab}),~~$
(\ref{ettta},\ref{etttb}) can be rewritten as
\be
\d x^a=M^k\l_{kl}\L^{al},~~\d t^a=mM^a.
\ee{ettu}
This has a formal resemblance with the following transformations
\be
\d x^a=\bar{M}\p^a,~~\d \p^*=m\bar{M},
\ee{ettv}
which are generated by
$p^a\p_a+m\p^*$ in the pseudo-classical model \cite{rgal}. This tells that
$M\tilde{\b}+L\tilde{\g}$ shows rough correspondence with
$\bar{M}(p^a\p_a+m\p^*)$ of the pseudo-classical model.
The lagrangian (\ref{ela}) is also very similar to the pseudo-classical
one
\be
\CL=p^a\dot{x}_a-N(p^2+m^2)
-\Ha\p^*\dot{\p}^*-\Ha\p^a\dot{\p}_a-\bar{M}(p^a\p_a+m\p^*),
\ee{susy}
upon the correspondences below:
\bE
\l^{ab}&\sim&\th^a\th^b\nn\\
\L^a{}_b\th^b&\sim&\p^a,\nn\\
\th^at_a&\sim&\p^*,\nn\\
M_k\th^k&\sim&\bar{M}.
\eE{eeep}
However, we cannot tell that correspondence in (\ref{ettt}), which may be
due to the different nature between boson and fermion. A further
investigation on the exact identification for those symmetries is to be
hoped.

\bigbreak\bigskip\bigskip\noindent{\bf Acknowledgments}

This work was supported in part by the Korean Science and Engineering
Foundation.


\end{document}